\documentclass[lettersize,journal]{IEEEtran}
\usepackage{amsmath,amsfonts}
\usepackage{algorithmic}
\usepackage{algorithm}
\usepackage{array}
\usepackage{textcomp}
\usepackage{stfloats}
\usepackage{url}
\usepackage{verbatim}

\usepackage{multirow}
\usepackage{graphicx}
\usepackage{cite}
\usepackage{color}
\usepackage{soul}
\usepackage{caption}
\usepackage{subcaption}
\usepackage{lipsum}
\begin{document}

\setlength{\textfloatsep}{3pt}
\setlength{\intextsep}{3pt}
\setlength{\floatsep}{3pt}

\title{Digital Network Twins for Next-Generation Wireless: Creation, Optimization, and Challenges}

\author{Zifan Zhang,
        Zhiyuan Peng, Hanzhi Yu, Mingzhe Chen~\IEEEmembership{Senior Member,~IEEE}, Yuchen Liu,~\IEEEmembership{Member,~IEEE}
}



\maketitle

\begin{abstract}
Digital network twins (DNTs), by representing a physical network using a virtual model, offer significant benefits such as streamlined network development, enhanced productivity, and cost reduction for next-generation (nextG) communication infrastructure. 
Existing works mainly describe the deployment of DNT technologies in various service sections.
The full life cycle of DNTs for telecommunication has not yet been comprehensively studied, particularly in the aspects of fine-grained creation, real-time adaptation, resource-efficient deployment, and security protection. This article presents an in-depth overview of DNTs, exploring their concrete integration into networks and communication, covering the fundamental designs, the emergent applications, and critical challenges in multiple dimensions. We also include two detailed case studies to 
illustrate how DNTs can be applied in real-world scenarios such as wireless traffic forecasting and edge caching. Additionally, a forward-looking vision of the research opportunities in tackling the challenges of DNTs is provided, aiming to fully maximize the benefits of DNTs in nextG networks.

\end{abstract}

\section{Introduction}

Although digital twin (DT) has been successfully applied in the fields of manufacturing and complex system operation and maintenance, research related to the creation, deployment, and application of DT for next-generation (nextG) networks is still in its infancy. 
In this regard, digital network twins (DNTs) emerge as a promising solution~\cite{ITUT2022Y3090}, offering a multitude of significant benefits, such as accelerated network development processes, enhanced resource utilization, easy management, and faster networking innovation with reduced costs. Specifically, in the network deployment phase, the availability of DNT can deliver a nextG network planning and operation tool to see the entire life cycle of network and management details before physically placing wireless sites. If system enhancement is desired, performance parameters can be adjusted and simulated in a DNT without imperiling the operation of the entire infrastructure. Then, during the network operation phase, DNT can enhance the access and analysis of already deployed network systems to enable remote commissioning and diagnostics, which lowers service costs by remotely configuring faulty parts of a network facility, e.g., ordered and replaced accordingly for the network operator or other involved researchers; Lastly, in the network design phase, a DNT can also be used to predict network performance and behavior in complex environments, design newer and better systems from the learned history of infrastructure operating conditions, and optimize facilities' efficiency and output.

However, before its full realization, representing a physical network using a virtual model faces many challenges: 1) how to build a high-fidelity twin involving various physical objects, communication entities, signal propagation phenomenon, and network attributes synergistically; 2) how to adaptively evolve the built DNTs taking into account the highly dynamic nature of nextG networks; 3)  how to integrate the DNT model with the existing network infrastructure and systems while considering communication and computational constraints; and 4) how to gather the representative data from a DNT and its physical counterparts for assisting various downstream applications.

To resolve these practical challenges, several surveys and tutorials about DNTs have been proposed~\cite{tao2024wireless,zhou2022network,hakiri2024comprehensive,tang2022survey,almasan2022digital,alcaraz2022digital,lin20236g, wen2022toward, han2022deliverable}, mainly describing the deployment of DNT technologies in various service sectors. 
\cite{wu2021digital,zhou2022network,hakiri2024comprehensive} present a broad survey to explore the integration of DTs into networks for potential application scenarios, such as aviation, healthcare, 6G networks, and intelligent transportation systems. \cite{tang2022survey} focuses on deploying DNTs at edge to improve performance, such as user throughput and security, and reducing the cost of communication, computation, and caching. \cite{almasan2022digital, lin20236g} outline the recent trends in machine learning and explore their integration into DNTs for efficient network management. \cite{alcaraz2022digital} analyzes the current state of the DNT paradigm and classifies the potential threats associated with it, considering its functionality layers and the operational requirements to ensure the trustworthy use of a DT. Specifically, \cite{wen2022toward} takes more complex information into account, capturing and representing network dynamics as well as interacting with the modeled reality in real-time. Nevertheless, despite the widespread interest in deploying DNTs for telecommunication, it has not yet been comprehensively studied about the full lifecycle of DNTs, particularly in fine-grained creation, real-time adaptation, resource deployment optimization, and security.

This article presents a broad overview of DNTs and unveils the full potential of their integration into communication networks, covering the fundamentals, the emergent applications, and critical challenges in multiple dimensions. 
We start by exploring challenges and opportunities in DNT model creation and adaptation, in the aspects of mapping solutions, twinning data generation, and real-time synchronization requirements. Then, we consider the joint optimization of mapping and resource management to deploy the DNTs in practical network systems, by integrating split learning, graph neural networks, and proximal policy optimization within a reinforcement learning framework. Moreover, the security issues within the DNTs are unveiled, with a particular focus on adversarial attacks and information leakage. Lastly, we showcase concrete application scenarios of DNTs supported by use-case experiments. The paper concludes by outlining future research directions in addressing the above technical challenges.

\begin{figure*}
	\centering
	\includegraphics[scale=0.72]{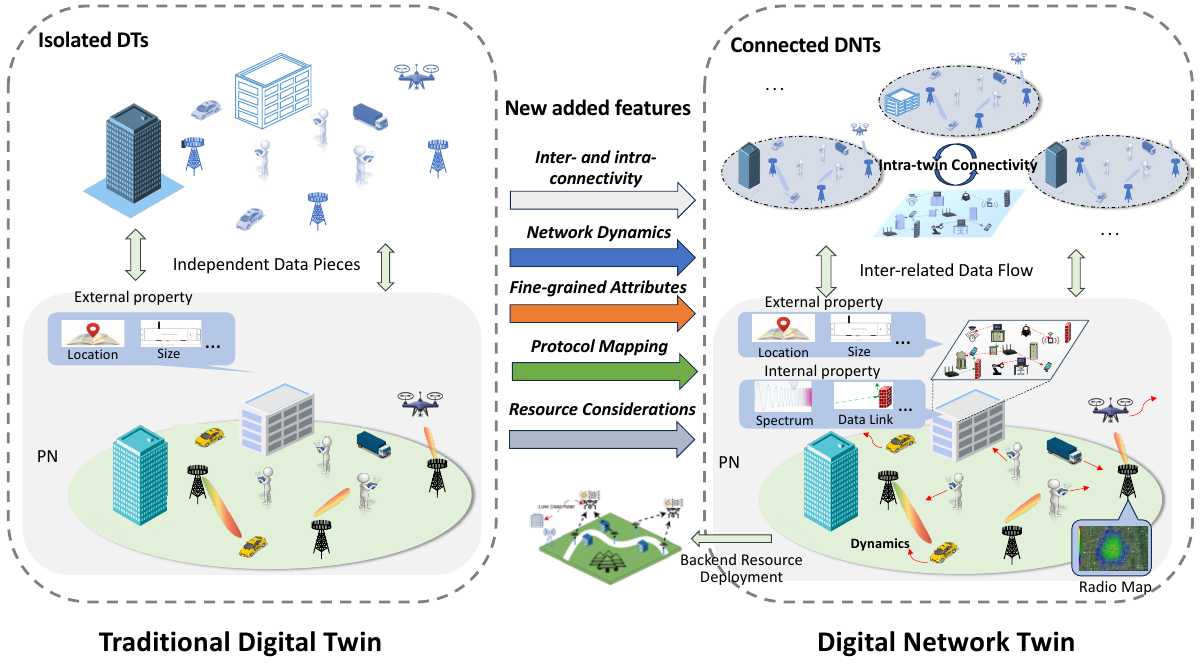}
	\caption{The differences between traditional DT and digital network twins (DNTs). DNTs go beyond by coordinating and managing the transitions between multiple DTs within a network to optimize overall performance in real time.
 }
	\label{fig:MVNarchitecture}
	\vspace{-3mm}
\end{figure*}

\section{Preliminaries and Overview}

A digital twin (DT) is a virtual representation of a physical system designed to reflect its runtime process and critical properties. It covers the entire lifecycle of each object, continuously updating from real-time data, and employs reasoning to predict future dynamics and inform decision-making. 
As an extension, DNTs offer advanced capabilities beyond traditional DTs by incorporating network-specific elements. 
While DTs enable continuous data exchange between isolated digital models and physical systems, DNTs extend this by jointly modeling, simulating, and optimizing complex communication networks in real time. 

As illustrated in Fig. \ref{fig:MVNarchitecture}, DNTs differ from traditional DTs by integrating both inter- and intra-connectivity information, including geological distance, coverage area overlap, and data distribution similarity.
Together with fine-grained network attributes among components within the physical network (PN), such as user traffic behaviors, the connected DNTs provide a comprehensive view of the entire system. 
In addition to capturing physical connectivity properties and network dynamics, DNTs maintain a synchronized mapping of the \textit{protocol stack states}—including PHY/MAC layer parameters (e.g., modulation scheme, channel coding, scheduling policies), transport configurations (e.g., congestion control settings), and application-level QoS requirements—alongside \textit{resource states} such as spectrum allocation, CPU/GPU utilization, and energy availability. This multi-layer state representation enables DNTs to reflect system-level configurations with high fidelity and temporal accuracy. Leveraging this holistic, cross-layer view, the network optimization problem is no longer solved by isolated DTs; instead, DNTs can coordinate state transitions among multiple twins and dynamically orchestrate resource allocation (e.g., bandwidth, computing cycles, storage) across interconnected twins. Such coordination is essential in large-scale, dynamic networks where traffic patterns, connectivity conditions, and service demands fluctuate.

Specifically, a PN consists of user devices (i.e., virtual reality headsets, sensors, servers, base stations (BSs), etc), network protocols, and communication channels. Specifically, a user device is defined as a physical network object (PNO) that can be described by a set of internal properties (i.e., physical dimension, location, mobility patterns) and a set of external properties (i.e., data transmission, access channel). In particular, external properties are used to capture how PNO affects other PNOs in the PN. On the other hand, the DNT, consisting of a network of twins, is a real-time virtual mapping of the PN such that it is expected to comprise the same components with the same internal and external properties.  
Specifically, the DNT is built based on the following two processes: 
1) Mapping all PNOs with their internal and external properties as well as their interrelations into a digital environment; 2) Self-evolving this digital environment according to the changes in any internal or external properties of PNOs in the PN. 
For example, a user device switches its data transmission from BS$_1$ to BS$_2$ in the PN, 
which will also affect the external properties (e.g., spectrum access) of other user devices due to the varying communication interference. Therefore, when the DNT updates the external property for a mobile device, it must also update the properties for other affected user devices accordingly.
This considered representation model will guide the fundamental designs of DNT creation, updates, and applications as discussed in the following sections.
Notably, while the existing works~\cite{wu2021digital,tang2022survey,almasan2022digital,alcaraz2022digital,lin20236g, wen2022toward} extensively explored the concept and applications of DNT architectures, such as simulating 5G networks and optimizing spectrum usage, practical challenges regarding efficient creation and real-time updates remain underdeveloped. 
For instance, a conceptual survey on DNTs in~\cite{wu2021digital} highlights how DTs enhance network performance through real-time monitoring and predictive analytics, emphasizing their role in intelligent network management. 
Another recent survey~\cite{alcaraz2022digital} explores the security landscape of DTs by categorizing potential threats and vulnerabilities in various downstream applications. It mainly emphasizes the increasing attack surface introduced by DTs, particularly due to their integration with physical and cyber domains. 
Orthogonal to existing literature, this article specifically addresses the foundational building blocks and online adaptation of DNTs within communication networks, introducing novel mapping mechanisms and resource deployment strategies to address a network of DTs.


\begin{table*}
\centering
\caption{Summary of the challenges and opportunities of the DNT.}
\label{fig:table}
\includegraphics[scale=0.4]{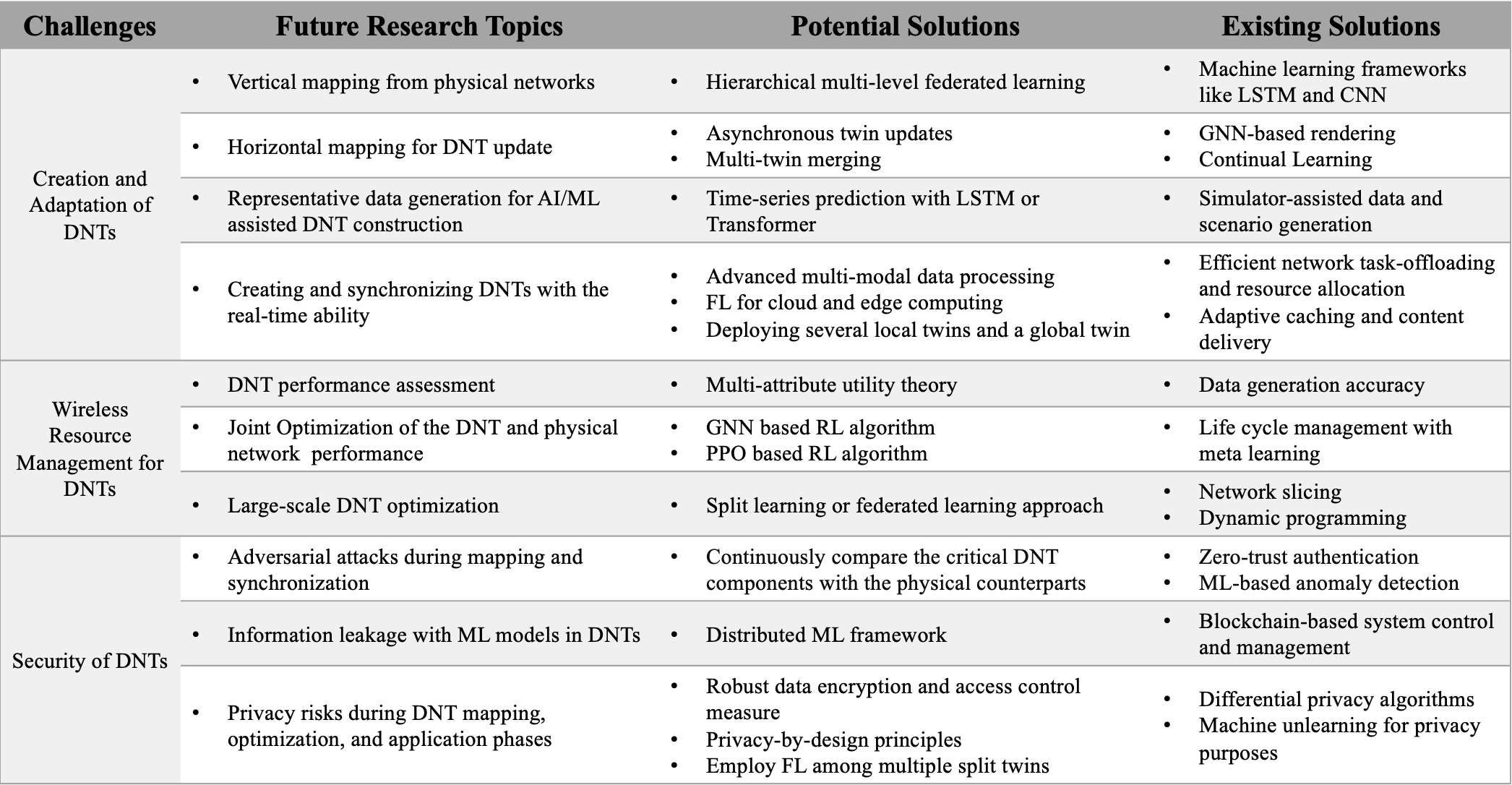}
\end{table*}


\section{Challenges and Opportunities}

This section addresses the challenges, mitigation strategies, and prospects of DNTs, as summarized in Fig.~\ref{fig:table}. We begin by examining the creation and adaptation of DNTs, with a particular emphasis on optimizing operational efficiency. Next, we explore their practical application in wireless resource management, where the continuous monitoring, assessment, and stabilization of DNT quality at scale pose significant challenges. Lastly, we highlight the security risks that emerge when integrating machine learning models into DNTs, especially in privacy-sensitive and mission-critical cyber infrastructures, acknowledging these as critical concerns.

\subsection{Creation and Adaptation of Digital Network Twin}\label{sec:dt_creation}

Recent studies typically employ machine learning (ML) methods~\cite{li2025generative}, such as LSTM and CNN, to build the initial DNTs. Combined with network simulators, these models create virtual scenarios that closely mirror the PN. However, However, these ML-based mapping process demands substantial communication and computation, as it typically requires offline training over large-scale datasets and potentially high-dimensional feature exchanges between distributed nodes, preventing them from supporting latency-sensitive services and continuous data synchronization.

%

\subsubsection{Vertical Mapping from Physical Networks}

One major evolution of nextG mobile and wireless networks is being driven by a move to higher carrier frequencies, e.g., millimeter-wave (mmWaves) and Terahertz (THz). Different from the lower-frequency radio, the diffraction ability of an mmWave/THz signal is much weaker and less reliable due to its smaller wavelength, which means that deployed objects in the environment have a significant impact on communication attributes and performance. This yields insight into deriving network attributes from environment details, including objects' locations, sizes, material types, and reflective properties. Thus, instead of cloning all details from the physical network to DNT, which is inefficient and even impractical, one possible solution is to map essential environment information from the physical networks, and then derive the corresponding attributes in the DNT locally through an environment-aware mapping approach as in~\cite{liu2022environment}. This necessitates employing machine learning frameworks like long short-term memory (LSTM) and convolutional neural networks (CNNs) to effectively capture spatial and temporal domain knowledge within a wireless network.
Synchronous federal learning (FL) is particularly effective for this vertical mapping because it requires all participating devices to update the global model simultaneously. This simultaneous updating process ensures high consistency across devices and improves overall model convergence, leading to more accurate and stable outcomes at this initial stage. Specifically, multiple twins can be created using a periodic clustering mechanism that virtualizes network regions covered by different base stations, taking into account their distinct geographical and wireless characteristics.

\vspace{+0.1cm}
\subsubsection{Horizontal Mapping for DNT update}
\begin{figure}
	\centering
	\includegraphics[scale=0.21]{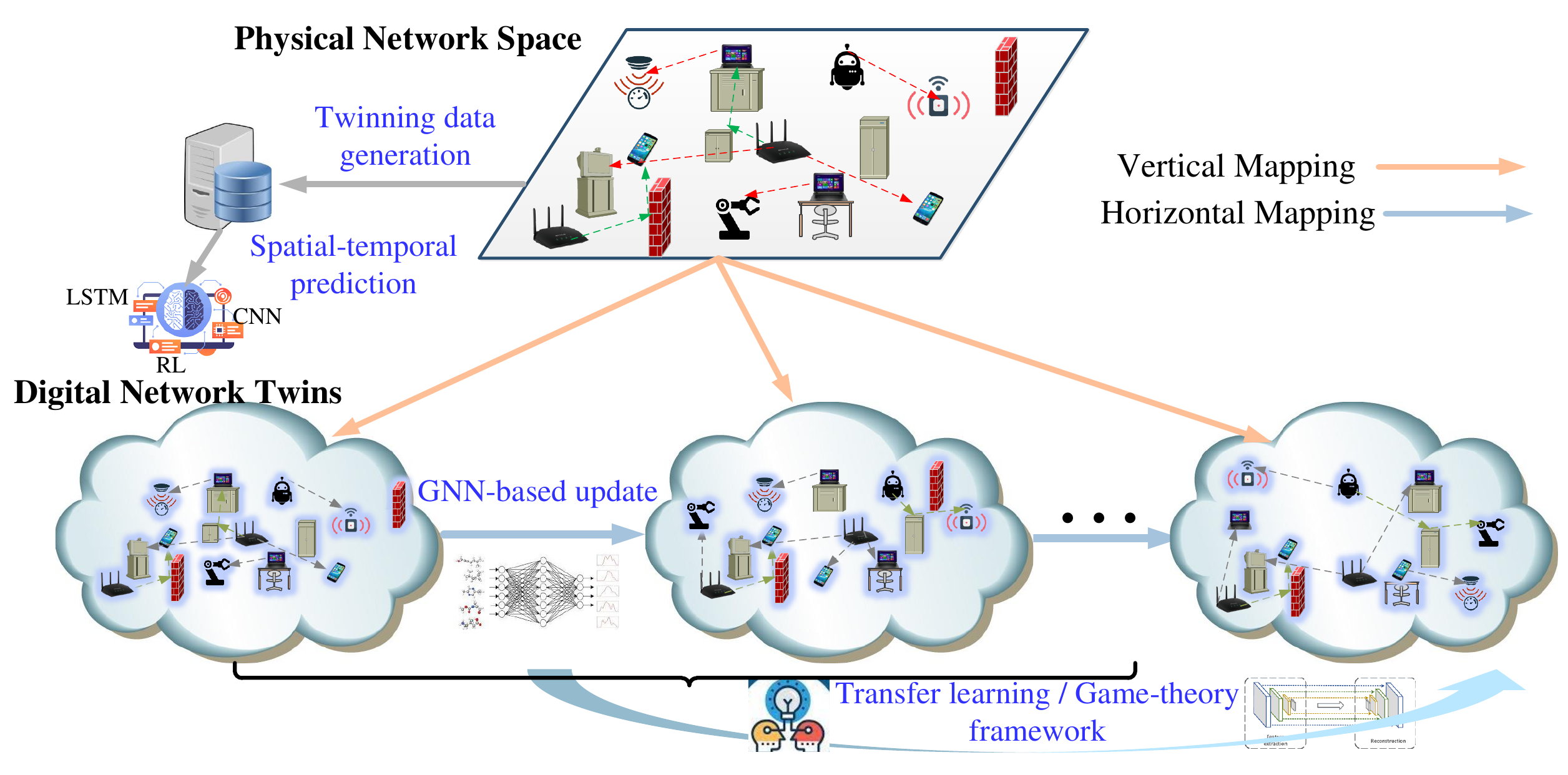}
	\caption{ML-assisted DNT mapping framework.}
	\label{fig:DNT_mapping}
	\vspace{-2mm}
\end{figure}

Different from the vertical mapping approach that gathers the information from physical environments to build a network twin, one primary goal of horizontal mapping is to reuse and migrate the existing DNTs and then create a new one when necessary (as shown in Fig.~\ref{fig:DNT_mapping} with the blue lines). 
To this end, two fundamental approaches can be exploited, including:
a) \textit{GNN-based rendering}: graph neural network (GNN) can be used to evolve existing DNT for environment dynamics. This includes detecting the variations of the physical network and then updating the DNT accordingly. Given the output of the GNN, an unsupervised network representation learning model \cite{zhang2018network} can be trained to determine whether to preserve the current DNT attributes and structure and then update the corresponding nodes according to the joint vertex embeddings when necessary. 
b) \textit{Multi-twin merging}: When mapping the physical network into the digital space, we may need to create multiple DNTs that fulfill different tasks via transfer learning and/or game theory, e.g., a \emph{radio-map twin} used for network planning and resource allocation, a \emph{channel-knowledge twin} for CSI acquisition and signal processing, etc. This involves the process of selecting appropriate twins among a set of available candidates, and then fusing the corresponding attributes in spatial and/or temporal domain.
Another objective is to continuously update the existing twin model. Asynchronous federated learning (FL) is particularly suited for this phase, as it permits devices to contribute updates to the global model at different times. This flexibility enhances scalability by allowing devices to participate despite varying availability, while also supporting different computational capacities, making the process more adaptable to diverse environments.

\vspace{+0.1cm}
\subsubsection{Twinning Data Generation}
One main challenge of utilizing AI/ML methods to construct DNTs is the necessity to generate a substantial volume of representative data capturing both environmental intricacies and coupled network attributes. However, this task is often impractical and unfeasible in real-world networks due to the extensive measurement campaigns required. Thus, one potential solution is to synthetically generate high-quality data covering a wide range of fine-grained network scenarios, which is then used to develop a faithful, efficient approach for the DNT creation. 
For instance, a quasi-deterministic (Q-D) ray tracer \cite{liu2022environment} can be used to generate various instance data within a known scenario covering different access points (AP)/user locations and protocol settings. 
Such a model can capture the geometrical properties of the channel for each transceiver and generate the profile of delay, path gain, angle of departure (AoD), angle of arrival (AOA), etc, for every path component.  
However, the data quality from the ray tracer is highly dependent on the abstracted channel model, which needs to be well-designed to acquire an accurate channel impulse response. In this regard, a small amount of measurement data from real-world experimental testbeds will contribute to the data generation process in two aspects: a) help derive the aligned scenario distributions, and b) create a feedback path to tune the related deterministic channel parameters such that the ray tracer can produce a more accurate dataset that matches the actual measurements.

\subsection{Wireless Resource Management for Digital Network Twins}  

Different from current resource management work that optimizes temporal and spatial allocations only for the physical network, resource management herein must consider networking resource (i.e., spectrum, transmit power, and computational power) management and mapping method selection to optimize the performance of both physical networks and DNTs. In particular, each mapping method requires unique physical network data to generate DNTs, thereby resulting in different data transmission requirements. For example, to map communication attributes between two entities, horizontal mappings require the digital counterpart to learn the network evolution, while vertical mappings mainly require environment parameters from the physical counterpart. In this regard, joint optimization of mapping method selection and resource management yields the following unique challenges. 

Current research treats DNTs mainly as testbeds for developing and evaluating existing solution space, with little focus on algorithms that optimize the DNTs’ own resource allocation or related services \cite{li2025generative}. For instance, reinforcement-learning policies trained to manage spectrum and computing resources in wireless networks do not transfer directly to DNT environments, where optimization objectives and constraints differ significantly due to limited exploration, inter-realm resource coupling, and non-stationary dynamics as the twin and real network co-evolve.

%

\vspace{+0.1cm}
\subsubsection{DNT Quality Assessment} To optimize the wireless resource management of DNT and the physical network, the first challenge is to design a novel performance metric that can capture how the mapping strategy, data collection, and transmission affect DNT creation and optimization over the actual network. Different from the current DT work that only maps physical objects into virtual space, DNT requires mapping network environment details and underlying network attributes into twins. Here, one promising method is to leverage multi-attribute utility theory -- a powerful multiple-criteria decision analysis tool that can effectively analyze multiple dependent and correlated variables in a joint utility function. Different from current multi-attribute utility functions that use weight parameters to summarize different metrics, which may not be applied for dependent metrics, multi-attribute utility theory builds a joint utility function that can capture the effects of all network attributes, protocols, and physical objects along the DNT creation, analogous to the generation of a joint probability distribution in probability theory, thereby enabling the analysis of the correlations among dependent network attributes and parameters. 

\vspace{+0.1cm}
\subsubsection{Optimized DNT Deployment} 

Another challenge is to address the optimization of the DNT performance (e.g., efficiency and accuracy of generation process) via determining mapping methods (i.e., horizontal or vertical mappings introduced in Subsection III~A) for each network entity and wireless resource (i.e., bandwidth, transmit power, beamforming parameters) for data transmission from physical networks to DNTs. Since mapping method selection and actual wireless resource allocation are always coupled and time-dependent, where the current mapping and resource allocation will explicitly affect future decisions, one may study the use of reinforcement learning (RL) to maximize the joint utility function derived from the quality assessment stage. 

However, implementing RL for joint mapping selection and wireless resource management encounters two main challenges. First, representing the dynamics of twining objects and network attributes as a state in an RL framework is non-trivial, as nextG networks may involve numerous object nodes with diverse potential dynamics, which cannot be adequately captured by a limited-size vector. One promising solution is to use a GNN framework with heterogeneous graphs to represent the network dynamics due to its ability to represent different physical networks and DNTs with a unique embedding vector and to analyze the inherent relationships among them. The second challenge 
lies in how to deal with the continuous action space over the actual network (e.g., with bandwidth and power allocation, beamforming process). This is because standard RL (e.g., Q-learning) that uses value functions to select optimal actions cannot deal with a continuous action space. While discretizing these optimization variables will sacrifice the DNT performance, this motivates the use of proximal policy optimization (PPO), which can stabilize and output an optimal action according to an RL state.

\vspace{+0.1cm}
\subsubsection{Large-scale Twin Optimization}

The training and implementation of a centralized RL framework for DNT optimization require the collection of all physical network dynamics and DNT states, thus probably leading to high communication overhead and implementation cost, particularly in a large-scale nextG wireless network. As a result, the RL model must be extended to a \emph{distributed} scheme, referred to as DRL, which enables distributed APs and user devices in a physical network to \emph{cooperatively} determine their mapping methods and resource utilization thus efficiently optimizing the utility function. To this end, one solution is to explore a split learning (SL) approach to decide on how to divide GNNs and RL models across different network entities, and how to leverage that division to share data and perform collaborative RL for the creation of DNTs. Unlike standard SL, which categorizes neural networks based solely on data distribution and learning performance, DNT uses separate neural network models that take into account factors such as computational resources, transmit power, wireless spectrum resources, and the dynamics of each device. Given the divided neural networks, the next step is to investigate different information-sharing procedures among DNTs and analyze their impacts on collaborative RL performance.

\subsection{Security of Inter-Realm Communication with DNTs}

While numerous state-of-the-art studies are centered on crafting and implementing DT systems across diverse domains, there remains a notable dearth of emphasis on safeguarding and enhancing the security and privacy aspects of these immersive, mixed realms. 
Specifically, the nascent DNTs and their interfaces to the physical network infrastructure might become potential targets for malicious activities. This underlines the emergence of less-explored security and privacy risks, among which are adversarial attacks that engender counterfeit representations, alongside the vulnerability of information leakage across distributed wireless devices and apps. 
The intricate interplay and added complexity of the DNT ecosystem give rise to less-recognized but impactful threats, especially in the following areas.

Prior work has concentrated on data-centric and hardware threats to DNTs—e.g., data poisoning and physical tampering \cite{alcaraz2022digital}. Broader risks like privacy risks, model poisoning, and backdoor attacks have received far less attention, but can degrade inter-realm system performance significantly.

\vspace{+0.1cm}
\subsubsection{Adversarial attacks during mapping and synchronization}
First, unauthorized data manipulation during the digital mapping process undermines trust in virtual services by empowering attackers to exploit manipulated digital models for deceptive services. Second, continuous synchronization can open doors to \textit{inter-realm adversarial attacks}, wherein subtle data perturbations may result in a persistent divergence between the digital and physical systems. Effectively countering data poisoning attacks throughout the digital mapping and synchronization process while preserving digital representation accuracy remains a challenging and under-explored task. To tackle this problem, potential solutions involve mitigating security risks posed by malicious actors attempting to manipulate the digital model at an early stage and establishing a new security framework against the inter-realm adversarial attack, such as continuously paring the critical digital network components with the physical counterparts to identify whether the mapping data are trustworthy or not, followed by a mask-predict based countermeasure to drop out or replace the poisoned data/model pieces while maintaining the freshness of digital representation.

\vspace{+0.1cm}
\subsubsection{Information leakage with ML models in DNTs}
Given that ML models form a core component of the DNTs, a significant threat model pertains to ML parameter leaks within the network. This involves potential adversaries like VR users and servers excluded from trustworthy training processes. These adversaries, whether active or passive, pose the risk of inferring raw data from both VR users and servers. The consequential risk is the potential leakage of sensitive information, encompassing user details as well as system-level operational details when leveraging these ML models in virtual environments.  One potential solution is to design a novel distributed learning framework that enables distributed wireless devices to collaboratively train ML models while preventing the leakage of sensitive data of connected users during the training of ML models for DNT generation and synchronization. This framework combines quantization theory and federated learning to enable wireless devices to efficiently generate a globally optimal ML model and prevent information leakage without sacrificing ML performance.

\vspace{+0.1cm}
\subsubsection{Privacy risks within DNTs}
The creation and use of DNTs raise significant privacy concerns due to the vast amount of data they capture, process, and store. These concerns encompass unauthorized access, misuse of sensitive information, surveillance, and data breaches. DNTs, being intricate and interconnected, risk inadvertently disclosing personal internet usage, preferences, and intellectual property to unauthorized parties. To address these concerns, robust data encryption and access control measures can be implemented to ensure secure and authorized access to the digital environment. Privacy-by-design principles can also mitigate risks by integrating privacy protections into the DNT mapping phase, including data minimization strategies during physical data collection, which balances the twin fidelity and system security. Additionally, employing techniques like decentralized federated learning among multiple split twins~\cite{stephanie2023digital} can enhance privacy by distributing model updates instead of raw data to a central physical network controller, and/or with differential privacy, adding noise to updates to prevent reverse engineering from DNTs to the physical networks.

\section{Case Studies}
This section showcases the creation and optimization of DNTs in three network case studies: wireless traffic forecasting, reliable edge caching, and secure DNT evolution, with respect to the proposed methods in Sec. III.
In these case studies, we explicitly address the four primary challenges outlined for constructing and deploying DNTs: \textit{scalability}, \textit{adaptability}, \textit{robustness}, and \textit{security}. First, the wireless-traffic forecasting scenario leverages hierarchical twin architectures (V-twin and H-twin) combined with both synchronous and asynchronous FL and strategic BS clustering, effectively addressing scalability and adaptability by significantly reducing modeling overhead and rapidly updating models in response to dynamic traffic conditions. Second, the reliable edge-caching scenario employs DNT-based environments that enable safe reinforcement learning by simulating rare and hazardous scenarios offline, thereby directly tackling the robustness challenge by minimizing the operational risk during policy training. Lastly, the case study on traffic poisoning attacks explicitly confronts the security challenge by introducing a Twin Inconsistency Defense (TID) scheme, which statistically trims anomalous model parameters to maintain attribute aggregation integrity under sophisticated attacks. Collectively, these case studies demonstrate practical strategies to address each challenge, highlighting the effectiveness and versatility of DNTs.

\subsection{Joint Vertical-Horizontal Twinning for Traffic Prediction}
In this section, we introduce the use of DNT for wireless traffic forecasting, where BSs can be clustered and virtually mapped into DNTs based on their distinct geological and network characteristics. As explained in Sec. \ref{sec:dt_creation}, these DNTs can be constructed by vertical mapping (\textit{V-twin}) and horizontal mapping (\textit{H-twin}) for DNT initialization and evolution, respectively. Here, V-twin exploits a synchronous federated learning framework to initialize a global twin model from the historical traffic data of virtualized network clusters. H-twin, on the other hand, is implemented with an asynchronous federated learning scheme that dynamically updates distributed twin models in response to network or environmental changes. The resulting self-evolved twins can then be applied for collaboratively forecasting precise time-series data, such as heterogeneous wireless traffic in cellular networks. 

\begin{figure}[h]
    \centering
    \includegraphics[scale = 0.3]{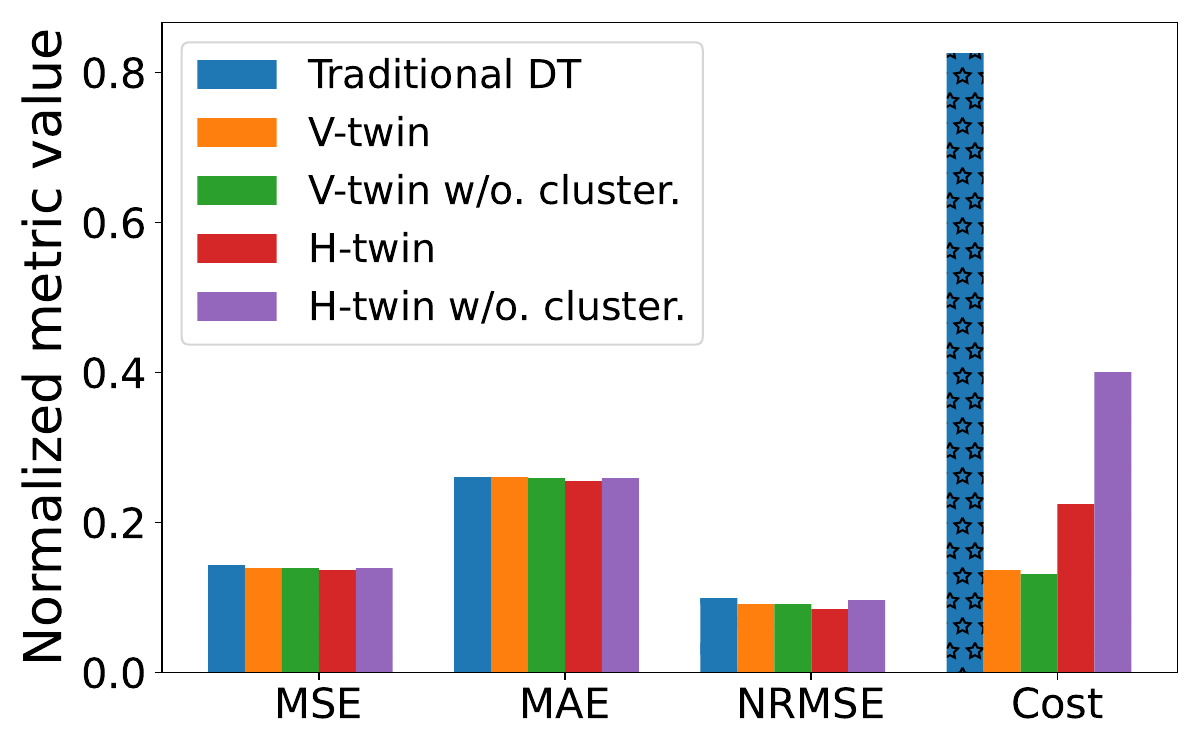}
    \caption{Mapping accuracy and deployment costs.}
    \label{fig:vh}
    \vspace{-2mm}
\end{figure}

Specifically, we employ the real-world network traffic dataset~\cite{barlacchi2015multi} of Milan City, which is composed of urban telecommunications behaviors intricately divided into 10,000 grid cells, with each cell equipped with a dedicated BS that captures a wealth of time-series data records, including detailed information on short message service (SMS), call activities, and internet usage patterns.
We adopt mean squared error (MSE), mean absolute error (MAE), and normalized root mean square error (NRMSE) as key performance metrics to measure the distinctions between ground-truth and generated data from the mapped DNTs. The ground truth data are time-series real-world traffic data collected from the physical cellular network.
Another key performance metric is the cost of the mapping process, which includes both training and communication time. In traditional DNT, all mapping is performed on a centralized server, requiring raw data to be uploaded for one-time modeling, which significantly increases deployment costs compared to our joint V-H twin method with continuous adjustments. H-twin incurs higher costs compared to V-twin, as DNTs demand continuous updates over extended periods rather than just initial construction. For comparison, we skip the BS clustering stage to create two twinning baselines, namely, \textit{V-twin w/o. cluster.} and \textit{H-twin w/o. cluster}. As shown in Fig.~\ref{fig:vh} with an example of SMS traffic data, all four twinning methods provide comparable prediction error rates, with the \textit{H-twin} achieving slightly lower prediction errors.
However, \textit{H-twin} needs to aggregate local twin models from the BSs and dispatch them back, thereby slightly increasing the communication cost. 

While VH-Twin offers only modest gains during the initial, one-off V-twin phase, its advantages become evident in the subsequent H-twin cycle. The horizontal stage delivers higher prediction accuracy than a \textit{Traditional DT} approach because asynchronous parameter updates assimilate the continuous, heterogeneous data stream, keeping the twin both stable and precise. It also cuts communication and computation costs by about 43.8\% relative to the baseline, single-level scheme. Taken together—slightly longer first-time initialization offset by far cheaper ongoing synchronization—the combined \textit{vertical-plus-horizontal} workflow provides a more resource-efficient solution for building and maintaining DNTs.

\subsection{DNT-Assisted Reliable Edge Caching}

DNTs can enhance edge caching optimization by embedding intervention modules into data-driven models for greater stability. Integrating safe RL algorithms with constraint modules, 
DNTs can be applied to optimize and safeguard caching decisions within the RL framework in a three-fold manner: 1) simulate network behavior and predict caching outcomes through accurate vertical mapping with historical data; 2) act as reliability nets, supporting risk-free exploration and policy testing in controlled environments (e.g., avoid overloading some BSs); and 3) generate rare events, allowing RL agents to train on sparsely distributed corner cases by producing synthetic yet realistic data across diverse conditions (e.g., maintain the hit rate in scenarios with highly dynamic content demand). 
\begin{figure}[h]
    \centering
    \includegraphics[scale = 0.35]{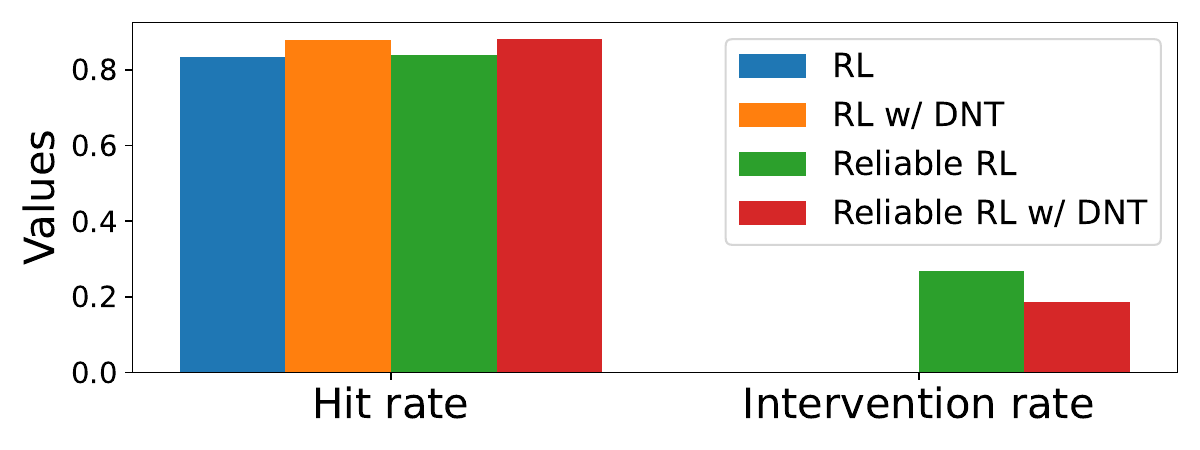}
    \caption{DNT facilitates reliable edge caching by accounting for reliability constraints and minimizing risks.}
    \label{fig:edge_cache}
    \vspace{-2mm}
\end{figure}
\begin{figure}[h]
    \centering
    \includegraphics[scale = 0.27]{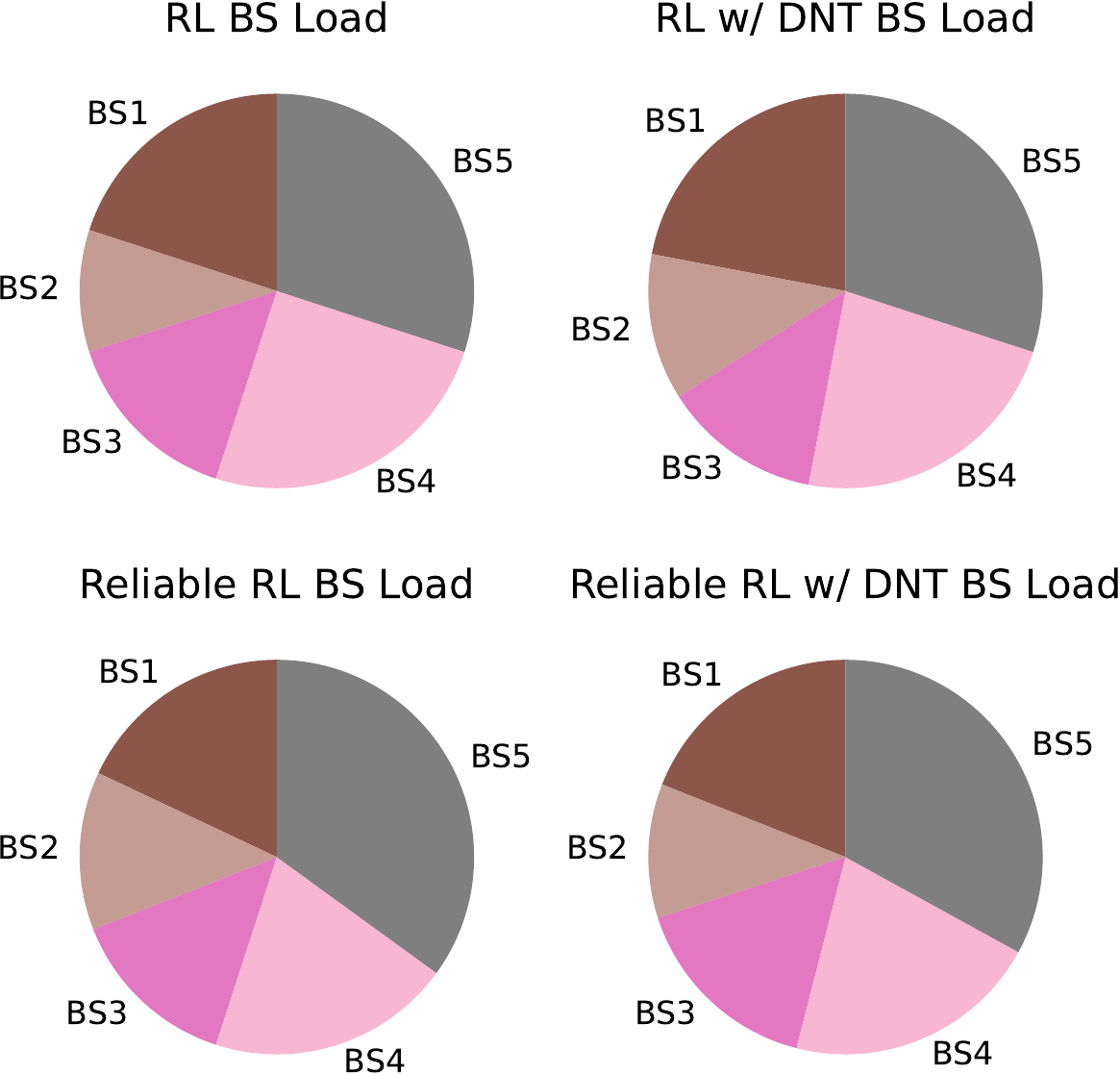}
    \caption{DNT with reliable RL balances loads among BSs.}
    \label{fig:edge_cache2}
    \vspace{-2mm}
\end{figure}
Here, we consider a cellular wireless network consisting of five BSs. Each BS is equipped with a local cache unit with a capability of 150 cache slots. A BS can only access its local cache unit and is forbidden to retrieve any cache units from its neighboring BSs. Each BS provides service for eight clients. The service can be overlapped, e.g., one client can be served by up to two BSs. We then create a DNT via vertical mapping from historical caching data, including information about user requests, contents, frequency, etc. The DNT serves as a data generator, providing ample datasets for training and predictive evaluation of different cache replacement policies. 
It produces one-step forecasts for the subsequently requested content. Moreover, various sets of historical caching data can be employed to feed the DNT, generating datasets that cover both common and rare wireless scenarios. 
We then implement four edge caching systems with or without the DNT. 1) \textit{RL}, a basic reinforcement learning model that uses data collected from physical cellular networks for training and prediction solely; 2) \textit{RL w/ DNT}, the model trained with data from both BSs and the virtual DNT; 3) \textit{Reliable RL}, the model augmented by integrating safety state, action and reward modules to monitor and reduce the load imbalance among BSs; 4) \textit{Reliable RL w/ DNT}, the reliable RL model trained with diverse data from both BSs and the virtual DNT. Figs.~\ref{fig:edge_cache} and \ref{fig:edge_cache2} evaluate the performance of four systems in terms of caching hit rate, number of risk interventions (introduced by safe modules from reliable RL), and BS load. The inclusion of
DNT with safety modules achieves a higher hit rate with balanced BS loads, hence enhancing the reliability and efficacy in managing cached content across densely deployed small-cell BSs. 

Conventional data-driven optimizers often overlook decision-making reliability, leaving BSs prone to overload, uneven traffic distribution, and poor quality of experience. D-REC remedies this gap by introducing control modules into a constrained Markov decision process. These modules continually reshape the action, reward, and state definitions to respect explicit reliability targets, thereby curbing the risk of large-scale failures. Here, DNTs act as RL co-processors and safety monitors, mining diverse traffic traces to forecast the impact of cache-replacement moves in dense small-cell deployments. The use of DNT also reduces the intervention from the elaborated safety constraints, suggesting that the augmented data from DNTs can improve the robustness of RL models with fewer interventions.

\subsection{Twin Poisoning Attacks to DNTs}

Accurate DNT application is increasingly critical as device counts and data-driven applications surge, yet distributed learning frameworks that build our DNTs are vulnerable to twin poisoning attacks. The attackers have the ability to subtly manipulate local twin model updates and further degrade network control and management. Existing defenses focus mainly on classification tasks, leaving regression-oriented DNT workloads, where adversaries manipulate continuous time-series inputs. We introduce an attack on distributed DNT-assisted wireless networks, namely Traffic Poisoning Injection (TPI), that forgoes hard-to-breach BSs by deploying low-cost fake DNTs. It uses only each attacker’s initial model and the current global DNT to construct fake but misleading local twin models to bypass the detection from the global DNT. To mitigate this threat, we present a Twin Inconsistency Defense (TID) strategy, which statistically trims dimension-wise outlier parameters and then re-weights the benign DNT models during the attribute aggregation process.

In this part, we also consider the same real-world network traffic prediction setting in Sec. IV-A. We benchmark our TPI attack against several state-of-the-art traffic poisoning attacks and defenses to underscore its effectiveness. Additionally, we employ these baseline attacks to demonstrate the efficacy of our TID defense strategy. TABLE II illustrates that TPI incapacitates every aggregation rule tested. For Mean and FLTrust, both MAE and MSE score over the 100-point cap in the construction (V-twin) and maintenance (H-twin) phases, meaning the predicted results deviate greatly from ground truth. Even Median, which is unaffected by the MPAF baseline attack, collapses under TPI, increasing from an MAE of 0.281 to 100.00 during V-twin and from 0.296 to 100.00 during H-twin. By contrast, MPAF degrades only Mean and FLTrust, leaving Median and TID essentially unchanged. Across all scenarios, TID is the only method that remains accurate. Both MAE and MSE remain at 0.281/0.107 during V-twin and 0.27/0.10 during H-twin under MPAF attacks, while TPI raises errors to 72.5/27.5, lower than the hard-failure level of competing rules. These results establish TPI as a potent, rule-agnostic poisoning strategy and TID as the most robust defense throughout the DNT lifecycle.

\begin{table}[ht]
\centering
\scriptsize
\caption{Performance on the Milan‐Internet Dataset for Both V-twin and H-twin.}
\label{tab:milan_combined}
\begin{tabular}{|c|c|ccc|ccc|}
\hline
\multirow{2}{*}{Defense} & \multirow{2}{*}{Metric} & \multicolumn{3}{c|}{V-twin} & \multicolumn{3}{c|}{H-twin} \\ \cline{3-8}
 &  & NO & MPAF & \textbf{FTI} & NO & MPAF & \textbf{TPI} \\ \hline
\multirow{2}{*}{Mean} 
 & MAE & 0.281 & 100.0 & 100.0 & 0.266 & 100.0 & \textbf{100.0} \\
 & MSE & 0.106 & 100.0 & 100.0 & 0.101 & 100.0 & \textbf{100.0} \\ \hline
\multirow{2}{*}{Median} 
 & MAE & 0.281 & 0.281 & 100.0 & 0.296 & 0.296 & \textbf{100.0} \\
 & MSE & 0.106 & 0.106 & 100.0 & 0.101 & 0.101 & \textbf{100.0} \\ \hline
\multirow{2}{*}{FLTrust} 
 & MAE & 0.312 & 100.0 & 100.0 & 0.297 & 100.0 & \textbf{100.0} \\
 & MSE & 0.114 & 100.0 & 100.0 & 0.109 & 100.0 & \textbf{100.0} \\ \hline
\multirow{2}{*}{\textbf{TID}} 
 & MAE & \textbf{0.281} & \textbf{0.281} & \textbf{72.453} & \textbf{0.266} & \textbf{0.266} & \textbf{72.458} \\
 & MSE & \textbf{0.106} & \textbf{0.107} & \textbf{27.548} & \textbf{0.101} & \textbf{0.102} & \textbf{27.543} \\ \hline
\end{tabular}
\end{table}

\subsection{Lessons Learned \& Discussion}

The first case study explores wireless traffic forecasting in a city-scale cellular network covering approximately 10,000 grid-based cells within Milan. Here, DNT functions primarily as a predictive analyzer, integrating both historical traffic patterns (V-twin) and incremental, localized updates via asynchronous federated learning (H-twin). Such scenarios, characterized by abundant spatial-temporal data and relatively low direct-testing risk, highlight DNTs' strengths in efficiently tracking and predicting dynamic network states, making them particularly suitable for network management and proactive resource provisioning tasks.
The second case study addresses reliable edge-caching in a small-cell cluster, consisting of five base stations with overlapping coverage and stringent QoS constraints. Here, the DNT acts as a critical sandbox and safety-net environment, synthesizing rare and potentially harmful network conditions to safely guide RL-based caching policies. Scenarios such as this, marked by high experimentation risks, tight latency requirements, and the need to safely explore counterfactual and rare-event scenarios, clearly benefit from the DNT’s capability to rigorously test policies offline prior to deployment.
In the third case study, we explore the vulnerability of distributed DNTs to model poisoning attacks within a real-world traffic prediction setting. This scenario emphasizes that applications involving decentralized learning with continuous and dynamic data inputs particularly benefit from robust anomaly-detection defenses embedded within the DNT framework.

\section{Conclusion and Future Directions}
This article presents a broad overview of DNTs and unveils the vast potential of their integration for nextG wireless networks. We start by introducing the basic concepts and fundamental design of DNTs, with three highlighted applications, including what-if analysis and network planning. We then describe the critical research challenges and opportunities of DNTs, in the aspects of creation, adaptation, optimization, and security. Lastly, we showcase the use of DNTs in two wireless scenarios: traffic forecasting and edge caching. We present a variety of research opportunities in addressing the challenges of DNTs, with the goal of fully harnessing their potential in future communication networks.

Looking forward, several promising research directions can enhance both the capabilities and practical deployment of DNTs. First, developing automated mechanisms for dynamic twin adjustment, where DNTs intelligently balance model fidelity and computational efficiency based on real-time network states, could further optimize predictive performance. Additionally, extending DNT architectures to seamlessly integrate heterogeneous network elements (e.g., satellite, drone-based, and terrestrial networks) will broaden their applicability, such as transferring existing DNTs to non-terrestrial network scenarios. From a security perspective, designing advanced, adaptive defenses against poisoning attacks for continuously streaming data remains critical. On the deployment front, creating standardized interfaces and integration frameworks for open RAN with xApps/rApps and core networks will facilitate broader adoption and interoperability of DNTs.

\bibliographystyle{IEEEtran}
\bibliography{refs}

\end{document}